\documentclass[conference]{IEEEtran}
\IEEEoverridecommandlockouts
\usepackage{cite}
\usepackage{amsmath,amssymb,amsfonts}
\usepackage{algorithmic}
\usepackage{graphicx}
\usepackage{textcomp}
\usepackage{xcolor} %highlight text

\usepackage{caption}  
\usepackage{graphicx}  
\usepackage{caption}
\usepackage{subfigure} 
\usepackage{verbatim}
\usepackage{soul}
\usepackage{booktabs}
\usepackage{makecell} 
\usepackage{booktabs} 
\usepackage{amsmath}
\usepackage{hyperref}

\def\BibTeX{{\rm B\kern-.05em{\sc i\kern-.025em b}\kern-.08em
    T\kern-.1667em\lower.7ex\hbox{E}\kern-.125emX}}
\begin{document}
\title{Sebica: Lightweight Spatial and Efficient Bidirectional Channel Attention Super Resolution Network}
\author{Chongxiao Liu\\liucx5266@gmail.com
}

\maketitle

\begin{abstract} 
Single Image Super-Resolution (SISR) is a vital technique for improving the visual quality of low-resolution images. While recent deep learning models have made significant advancements in SISR, they often encounter computational challenges that hinder their deployment in resource-limited or time-sensitive environments. To overcome these issues, we present Sebica, a lightweight network that incorporates spatial and efficient bidirectional channel attention mechanisms. Sebica significantly reduces computational costs while maintaining high reconstruction quality, achieving PSNR/SSIM scores of 28.29/0.7976 and 30.18/0.8330 on the Div2K and Flickr2K datasets, respectively. These results surpass most baseline lightweight models and are comparable to the highest-performing model, but with only 17\% and 15\% of the parameters and GFLOPs. Additionally, our small version of Sebica has only 7.9K parameters and 0.41 GFLOPS, representing just 3\% of the parameters and GFLOPs of the highest-performing model, while still achieving PSNR and SSIM metrics of 28.12/0.7931 and 0.3009/0.8317, on the Flickr2K dataset respectively. Additionally, Sebica demonstrates significant improvements in real-world applications, specifically in object detection tasks, where it enhances detection accuracy in traffic video scenarios.
\end{abstract}

\begin{IEEEkeywords}
Super Resolution, Lightweight, Spatial Attention, Efficient Bidirectional Channel Attention
%Enter keywords or phrases in alphabetical 
%order, separated by commas. For a list of suggested keywords, send a blank 
%e-mail to keywords@ieee.org or visit \underline
%{http://www.ieee.org/organizations/pubs/ani\_prod/keywrd98.txt}
\end{IEEEkeywords}

\section{Introduction}
Single Image Super-Resolution (SISR) is a vital image processing technique that enhances the visual quality of low-resolution (LR) images by converting them into high-resolution (HR) versions with increased detail and pixel density. This technique finds extensive application in fields such as remote sensing, medical imaging, and multimedia enhancement. However, the process of converting LR images to HR faces significant challenges due to the multitude of potential high-resolution outcomes. While traditional methods, including interpolation and representation-based techniques, have been used to address these issues, recent advancements in deep learning, especially with models like the Super-Resolution Convolutional Neural Network (SRCNN), have markedly improved performance. Despite these advancements, most SISR models require powerful GPUs and involve complex networks with numerous parameters, which makes tasks that are sensitive to speed, such as real-time processing of video, challenging. Therefore, there is a growing need for more lightweight and computationally efficient Super Resolution models.

\begin{figure}[htb]
\centering
\includegraphics[width=1.0\columnwidth]{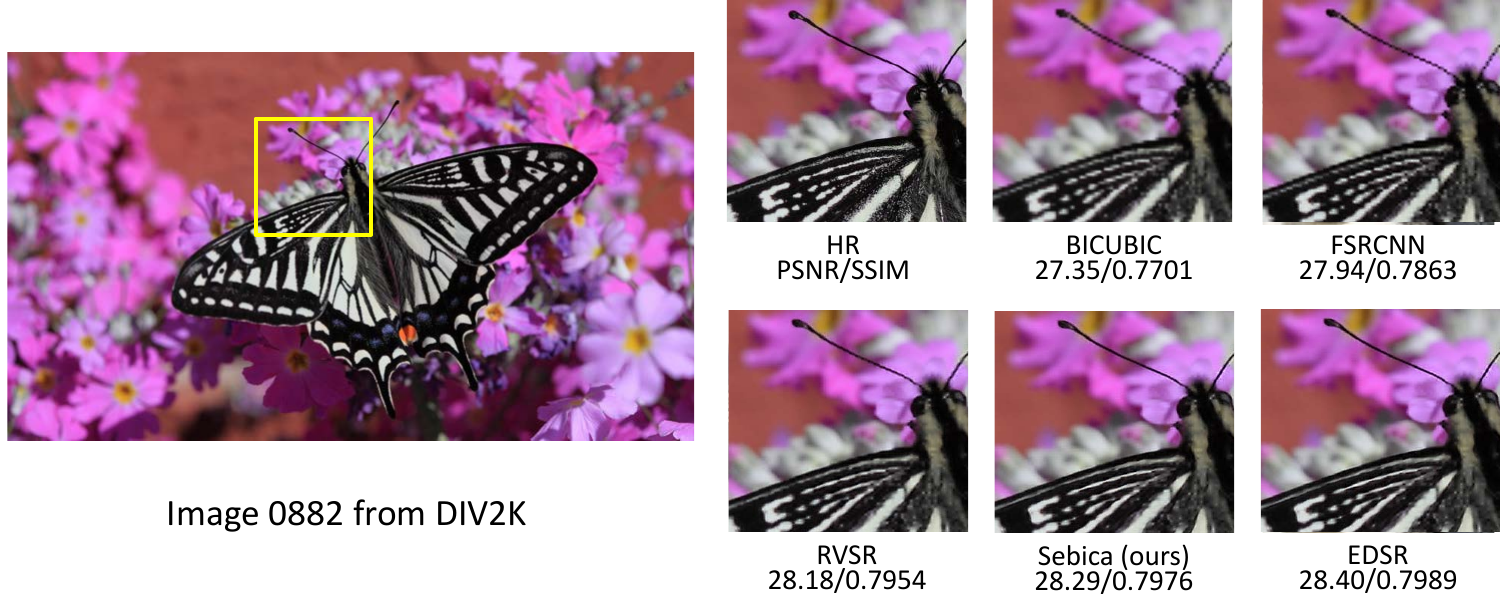}
\caption{\label{fig:comparison_1} 
Comparison of the SR images derived from listed networks.
}
\end{figure} 
Recently, various state-of-the-art networks have utilized attention mechanisms, leading to notable improvements and refinements in the reconstruction of high-quality HR images with low latency.
The lightweight nature and simplicity of these implementations have made it easier to integrate them directly into the feature extraction process of convolutional neural networks. This motivates us to incorporate attention mechanisms into SR, particularly for fast applications.

In this study, we present a lightweight Spatial and Efficient Bidirectional Channel Attention Super Resolution network, named \textbf{Sebica}, specifically designed to improve inference speed while balancing the trade-off in image quality during high-resolution conversion. The network incorporates multiple attention blocks that combine spatial and channel attention, working in conjunction with bilinear interpolation to upsample high-resolution images. In the spatial attention module, we use channel-wise average and maximum computations instead of traditional pooling methods, while introducing a bidirectional mechanism in the channel attention module. To enhance efficiency, 1D convolutions are employed to reduce computational costs. This approach dynamically adjusts the importance of channels, improving feature selectivity. Additionally, residual connections are used throughout the network to stabilize training and retain crucial information. Finally, the PixelShuffle operation is applied in the tail module to efficiently manage image upscaling.

We evaluate Sebica's reconstruction quality using the Peak Signal-to-Noise Ratio (PSNR) and Structural Similarity Index (SSIM) metrics on the KIV2K and Flickr datasets, as shown in Figure~\ref{fig:comparison_1}. The figure demonstrates that Sebica effectively reconstructs low-resolution images. Furthermore, we conduct experiments by deploying it in a traffic video object detection task, measuring its effectiveness compared to scenarios without Sebica's enhancement using metric of mean average precision (mAP).

The rest of the paper is organized as follows. Related work is discussed in Section~\ref{sec:related}, where we provide background information on super-resolution and review lightweight versions of these techniques. In Section~\ref{sec:method}, we detail our approach and the network structure of Sebica, with a specific focus on the attention mechanisms used to optimize the network. In Section~\ref{sec:eval}, we outline our evaluation steps, including the dataset and testbed setup. Finally, Section~\ref{sec:conc} concludes the paper and discusses future work. Our source code is available at: 
\href{https://github.com/idiosyncracies/Sebica}{https://github.com/idiosyncracies/Sebica}

\section{Related Work}\label{sec:related}

Early approaches to SR relied on techniques such as bicubic interpolation. Nowaday, the advent of deep learning revolutionized the field.  Dong et al. pioneered the use of deep convolutional neural networks (CNNs) in SR with SRCNN~\cite{dong2015image}, a three-layer model that significantly outperformed traditional methods. Later, deeper networks like VDSR~\cite{kim2016accurate}, EDSR~\cite{lim2017enhanced}, and RCAN~\cite{zhang2018image} achieved further improvements by leveraging deeper architectures and advanced residual blocks. Increasing the depth and width of CNNs has been effective in enhancing super-resolution (SR) performance, but it also significantly increases computational demands, making these models less suitable for lightweight environments such as mobile or edge devices, or for low-latency applications like real-time 4K video streaming. This challenge has led to the concept of lightweight and efficient super-resolution (LESR). Recent research has focused on designing more efficient architectures to meet these requirements.

LESR models have been developed to reduce computational complexity while maintaining high-quality image reconstruction. One prominent approach is FSRCNN~\cite{dong2016accelerating}, which extracts features in a low-dimensional space to minimize the computational cost and utilizes transposed convolution for upscaling. DRRN~\cite{tai2017image} and DRCN~\cite{kim2016deeply} introduced recursive mechanisms to share parameters and reduce the model size, while CARN~\cite{ahn2018fast} used group convolutions and cascading residual blocks to balance HR conversion quality and efficiency. Other techniques, such as the information distillation network (IDN)~\cite{hui2018fast}, further optimize efficiency by splitting feature maps and applying convolutions only to a portion of the data, reducing redundancy. Methods like IMDN~\cite{hui2018fast}, and then residual feature distillation block RFDB~\cite{liu2020residual}, which built on IMDN, who won AIM 2020~\cite{zhang2020aim}.

Moreover, compression techniques have also proven effective in improving efficiency. Approaches such as network pruning \cite{zhou2020rethinking, chao2020directional, zhan2021achieving}, knowledge distillation \cite{hui2018fast, liu2020residual, yu2023dipnet, angarano2023generative, xie2023large}, and re-parameterization \cite{wang2022repsr, gao2022rcbsr} have successfully accelerated high-resolution (HR) reconstruction. For example, RVSR leverages re-parameterization technique to support real-time 4K video processing and was highlighted at AIS 2024 \cite{conde2024real}. However, these methods often deliver suboptimal performance due to rigid schedules or inflexible modules. Additionally, lightweight models such as IMDN \cite{hui2018fast} and RFDB \cite{liu2020residual} fail to address inter-channel redundancy effectively, resulting in unnecessary computations, which can negatively impact robustness and generalization.

To address these challenges and enhance computational efficiency without sacrificing robustness across diverse applications, we propose Sebica. Instead of relying on compression techniques, Sebica refines the super-resolution (SR) network architecture by integrating a spatial and efficient bidirectional channel attention mechanism, along with 1D convolution to reduce computational overhead.

In the next section, we provide a detailed discussion of these features.

\section{Method}
\label{sec:method}

This section explores the details of the proposed method, beginning with an overview of the overall architecture. Following this, we detail the individual components, namely the spatial attention block and the efficient bidirectional channel attention module. Lastly, we present the loss functions utilized in the method.
\subsection{Overall Framework}
\begin{figure*}[htb]
\centering
\includegraphics[width=1.6\columnwidth]{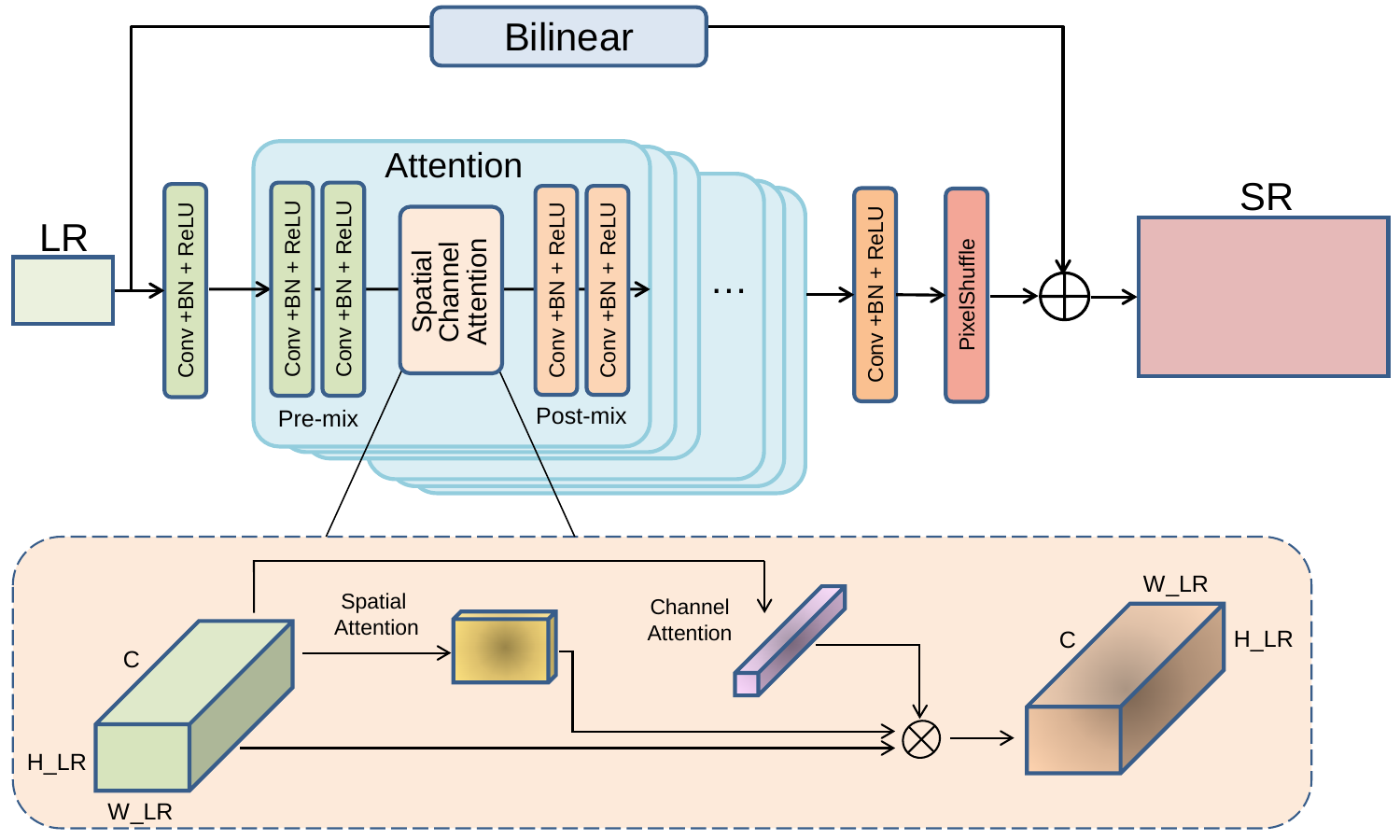}
\caption{\label{fig:overview} 
The attention module comprises 6 attention blocks, each integrating a spatial and an efficient bidirectional channel attention mechanism. Additionally, we propose a smaller version called Sebica\_small, which consists of 4 attention blocks. Spatial attention and channel attention mechanism are discussed in Section~\ref{sec:spatial_attn} and Section~\ref{sec:channel_attn} separately.}
.
\label{framework}
\end{figure*}

The overall network structure of Sebica is depicted in Figure~\ref{fig:overview}. The pipeline consists of a spatial and channel integrated attention module, convolution layers, pixel shuffle layers, and a bilinear mechanism.
The core component is the attention module, which integrates spatial and efficient bidirectional channel attention mechanisms. The spatial attention blocks capture pixel-wise relationships within the spatial dimensions of the feature maps, while the bidirectional channel attention modules effectively gather attention information across the channels derived from these feature maps, taking into account both forward and backward dependencies.
We will discuss this in more detail in Section~\ref{sec:attn}. 

\subsection{Attention module}
\label{sec:attn}

\subsubsection{Spatial Attention}
\label{sec:spatial_attn}
The spatial attention blocks capture detailed insights from the feature maps at the pixel level. Figure~\ref{fig:spatial_atention} illustrates the steps of the spatial attention mechanism for each component. The input is the feature map pre-mixed by the previous step, with dimensions of $C \times H \times W$,  where $C$ is the number of channels, $H$ is the height of the low-resolution (LR) image, and $W$ is the width of the LR image. Instead of using standard pooling operations, we employ channel-wise maximum and average pooling to simplify the process, which has been experimentally verified to not result in any loss of accuracy. The outputs from these pooling operations are then concatenated to form a new feature map, allowing us to capture both the maximum and average textures. This concatenated feature map is normalized using the sigmoid function and further processed through a two-dimensional convolution to convert its shape at $1 \times H \times W$. This processed feature map is later applied through element-wise multiplication with the input feature map being attended, after being normalized again with the sigmoid function.

The formulations for each step are presented below:   

\begin{figure}[htb]
\centering
\includegraphics[width=1\columnwidth]{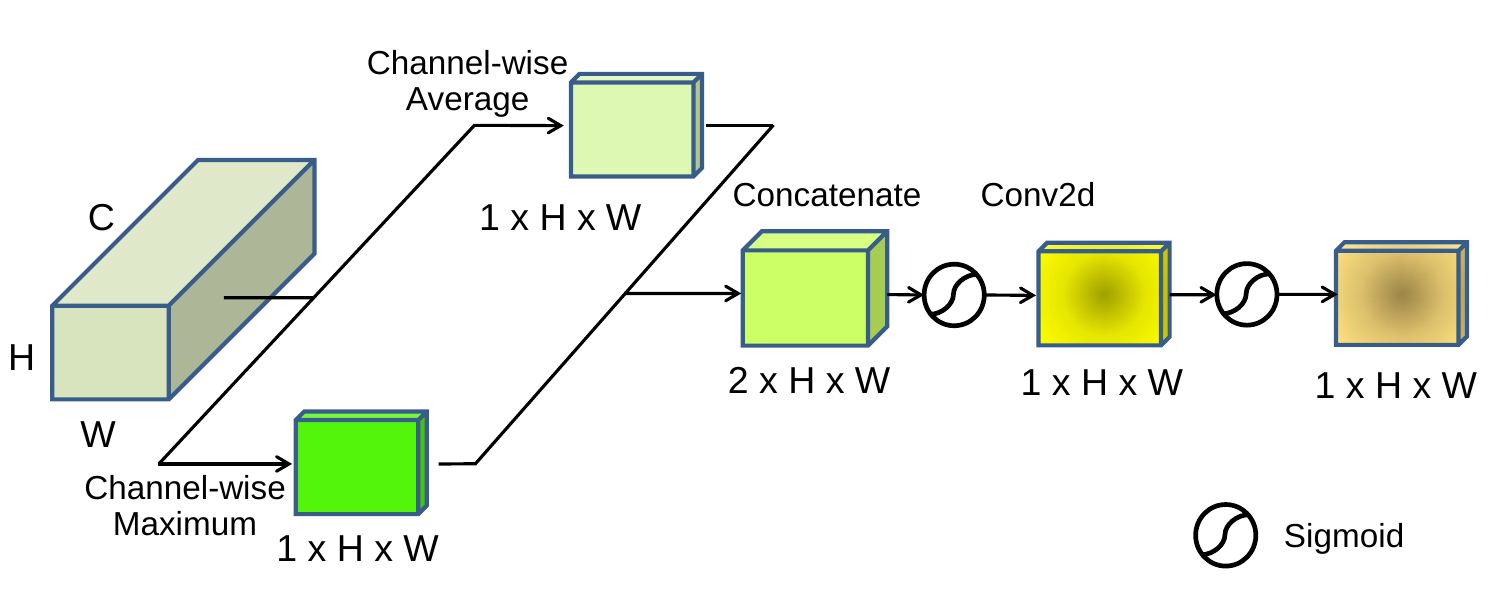}
\caption{ 
Spatial  attention block. To minimize computation while maintaining performance, we use dimension-wise average and max operations instead of traditional average and max pooling.
}
\label{fig:spatial_atention}
\end{figure}

\begin{itemize}
\item Compute Channel-wise Maximum and 
Average:

We apply two types of operations on the feature maps across channels, which are illustrated in Equation~\ref{eq:av_max_channel}, where $X_{c,i,j}$ represents the feature map at channel $c$, and spatial position $(i,j)$.

\end{itemize}

\begin{equation}
\label{eq:av_max_channel}
\begin{aligned}
    \mathbf{F_{\text{avg}}}(i, j) &= \frac{1}{C} \sum_{c=1}^{C} X_{c, i, j} \\
    \mathbf{F_{\text{max}}}(i, j) &= \max_{c=1}^{C} X_{c, i, j}
\end{aligned}
\end{equation}

\begin{itemize}
\item Concatenation of Average and Maximum Maps:

We concatenate the two feature maps $\mathbf{F_{\text{avg}}}$ and $\mathbf{F_{\text{max}}}$ along the channel dimension to form a 2-channel tensor of shape (2, H, W), as show in Equation~\ref{eq:spatial_attn_concat}.
\begin{equation}
\label{eq:spatial_attn_concat}
    \mathbf{F_{\text{concat}}} = concat(\mathbf{F_{\text{avg}}}, \mathbf{F_{\text{max}}})
\end{equation}

\item Convolution for Spatial Attention:

We apply a convolutional layer with a kernel size of 7$\times$7 to the concatenated feature map. This helps capture spatial dependencies in a local region. The convolution transforms the 2-channel input (2, H, W) into a 1-channel output (1, H, W). The calculation is illustrated in Equation~\ref{eq:spatial_attn_conv}, 
where $\mathbf{W_{\text{s}}}$ is the convolution filter, and $\sigma$ is the sigmoid activation function to generate the attention weights.

\begin{equation}
\label{eq:spatial_attn_conv}
    \mathbf{F_{\text{spatial}}} = \sigma(\mathbf{W_{\text{s}}} * \mathbf{F_{\text{concat}}})
\end{equation}

\item Applying Spatial Attention

We perform element-wise multiplication between the spatial attention map, with a shape of (1, H, W), and the input feature map to selectively enhance or suppress specific spatial locations, as show in Equation~\ref{eq:spatial_attn}, where $X$ is the original input feature map.
\begin{equation}
\label{eq:spatial_attn}
    \mathbf{A_{\text{spatial}}} = X\times \mathbf{F_{\text{spatial}}}
\end{equation}

\end{itemize}

\subsubsection{Channel Attention}
\label{sec:channel_attn}
\begin{figure}[htb]
\centering
\includegraphics[width=1\columnwidth]{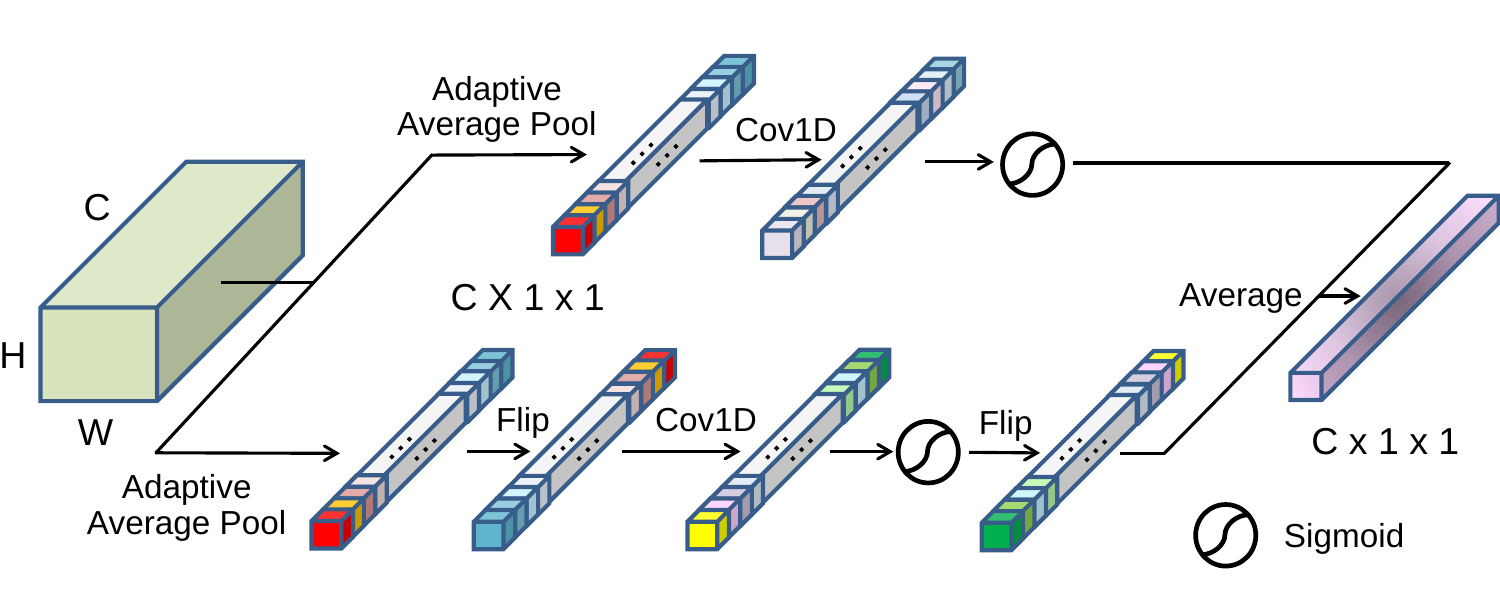}
\caption{ 
Efficient bidirectional channel attention block. Each cubic in various color represents one channel. Standard Sebica configures total 16 channels, we setup 8 channels for Sebica\_small version.}
\label{fig:channel_atention}
\end{figure}

Inspired by the success of Squeeze-and-Excitation (SE)~\cite{hu2018squeeze}  and ECA (Efficient Channel Attention)~\cite{wang2020eca} mechanisms in recent studies, which effectively use 1D convolution to learn inter-channel insights, we design an efficient channel attention block that further applies both forward and backward propagation. This block, called efficient bidirectional channel attention, enhances the extraction of channel significance from feature maps while maintaining a lightweight structure. As depicted in Figure.~\ref{fig:channel_atention}, the proposed block includes both conv\_forward and conv\_backward` to process information in forward and backward directions, respectively. During the forward pass, both the normal forward convolution and the backward convolution (achieved by flipping the input tensor along the last dimension) are performed. The results from both directions are then combined and passed through a sigmoid function to generate the final attention map.

The steps along with the equations illustrate as follows:
\begin{itemize}
 \item  Adaptive Average Pooling:

Adaptive Average Pooling dynamically adjusts the pooling window size to produce an output feature map of a specified size, allowing for flexible and size-invariant feature extraction.
We apply Adaptive Average Pooling for both forward and backward attention. For an input tensor $X \in \mathbb{R}^{C \times H \times W}$ , where C is the number of Channels, 
H is the height, and W is the width, Equation~\ref{eq:adaptive_ave_poolingt} indicates the Adaptive Average Pooling process:

Given an input feature map \( X \) of shape \( (C, H_{\text{in}}, W_{\text{in}}) \) and a desired output size \( (H_{\text{out}}, W_{\text{out}}) \):

\begin{equation}
\label{eq:adaptive_ave_poolingt}
F[:, h, w] = \frac{1}{h_{\text{pool}} \cdot w_{\text{pool}}} \sum_{i=h \cdot h_{\text{step}}}^{(h+1) \cdot h_{\text{step}}-1} \sum_{j=w \cdot w_{\text{step}}}^{(w+1) \cdot w_{\text{step}}-1} X[:, i, j]
\end{equation}

where: \\
\( h_{\text{pool}} = \left\lceil \frac{H_{\text{in}}}{H_{\text{out}}} \right\rceil \) \\
\( w_{\text{pool}} = \left\lceil \frac{W_{\text{in}}}{W_{\text{out}}} \right\rceil \) \\
\( h_{\text{step}} = \left\lfloor \frac{H_{\text{in}}}{H_{\text{out}}} \right\rfloor \)
\( w_{\text{step}} = \left\lfloor \frac{W_{\text{in}}}{W_{\text{out}}} \right\rfloor \)\\

$h$ and $w$ are the indices of the output positions, 
$h_{\text{pool}}$ and $w_{\text{pool}}$ are the heights and widths of the pooling windows, and $h_{\text{step}}$ and $w_{\text{step}}$ are the strides for the pooling operation.

% Global Average Pooling
% \begin{equation}
% \label{eq:global_ave_poolingt}
% \mathbf{F} = \frac{1}{H \times W} \sum_{i=1}^{H} \sum_{j=1}^{W} X_{i,j}
% \end{equation}
% \item Forward Channel Attention

\item 1D Convolution:

We apply 1D convolution, using the kernel in size of 3, to the pooled feature map. This convolution operates on the channel dimension to learn the dependencies among channels, where input feature map is denoted as 
$ F \in \mathbb{R}^{ C \times 1 \times 1}$, and output feature map is denoted as 
$ F_{\text{forward}} \in \mathbb{R}^{C \times 1 \times 1}$, hence output feature maps of forward by 1D convolution are denoted as Equation~\ref{eq:forward_1d_conv}.

\begin{equation}
\label{eq:forward_1d_conv}
\mathbf{F_{\text{forward}}} = \text{Conv1D}(\mathbf{F}) 
\end{equation}

\item Backward Channel Attention:

In backward channel attention, we first reverse the order of the channels in the feature map pooled with the Equation~\ref{eq:adaptive_ave_poolingt}, using a flip operation and then apply a 1D convolution. This enables the model to learn dependencies from a reversed channel perspective. Equation~\ref{eq:backward_1d_conv} presents the calculation, where $F_{\text{forward}} \in \mathbb{R}^{C \times 1 \times 1}$ is the backward attention map, and $\sigma$ is the sigmoid activation function to normalize the output between 0 and 1.
\begin{equation}
\label{eq:backward_1d_conv}
\mathbf{F_{\text{backward}}} = \sigma(\text{Conv1D}(\text{flip}(\mathbf{F})))
\end{equation}

\item Combining Forward and Backward Attention:

After obtaining the forward and backward attention maps, we combine them by averaging the two attention scores, ensuring that both directions contribute equally to the final attention map. The formulation is denoted in Equation~\ref{eq:bi_directional_attn}, where $A_{\text{channel}} \in \mathbb{R}^{C \times 1 \times 1}$ is the final combined channel attention feature map.

% Combined Channel Attention
\begin{equation}
\label{eq:bi_directional_attn}
\mathbf{A_{\text{channel}}} = \frac{\mathbf{F_{\text{forward}}} + \text{flip}(\mathbf{F_{\text{backward}}})}{2}
\end{equation}

\end{itemize}

\subsection{Loss Function}

We use Peak Signal-to-Noise Ratio (PSNR) as the loss function, which is derived from Mean Squared Error (MSE).

The PSNR loss function can be defined as follows:

\begin{equation}
\begin{aligned}
\text{MSE} &= \frac{1}{HW} \sum_{i=1}^{H} \sum_{j=1}^{W} \left( X(i,j) - \hat{X}(i,j) \right)^2 \\
\text{PSNR} &= 10 \cdot \log_{10} \left( \frac{\text{$max\_val$}^2}{\text{MSE}} \right)
\end{aligned}
\end{equation}

where $X(i,j)$ is the target tensor, $\hat{X}(i,j)$ is the output tensor, and $max\_val$ is the maximum possible pixel value.

The PSNR loss function is defined in Equation~\ref{eq:loss}.

\begin{equation}
\text{Loss} = \frac{1}{\text{PSNR} + \epsilon}
\label{eq:loss}
\end{equation}

When MSE is zero, the Loss is also zero. To address potential issues when \(\text{PSNR}\) is equal to zero, we introduce a small constant \(\epsilon\). Thus, the complete loss function becomes:

\begin{equation}
\text{Loss} = \begin{cases} 
0 & \text{if } \text{MSE} = 0 \\
\frac{1}{10 \cdot \log_{10} \left( \frac{\text{max\_val}^2}{\text{MSE}} \right) + \epsilon} & \text{otherwise}
\end{cases}
\label{eq:complete_loss}
\end{equation}
\\

\section{Ealuation}\label{sec:eval}

We use DIV2K~\cite{agustsson2017ntire} and Flickr2K~\cite{timofte2017ntire} datasets to train and evaluate our proposed network on one commodity computer as the test-bed. 
We evaluate accuracy using the PSNR and SSIM metrics, comparing our results with state-of-the-art lightweight super-resolution (SR) models. Additionally, to assess the effectiveness of our approach, we utilize Sebica for object detection on a publicly available video dataset \cite{car_object_detection}. This allows us to compare object detection performance, both with and without the application of super-resolution from our proposed network, using mean Average Precision (mAP) as the evaluation metric.

\subsection{Experiment Setup}
\label{sec:hw}
\begin{itemize}
\item Datasets:

The DIV2K dataset consists of 1,000 high-resolution images, divided into 800 training, 100 validation, and 100 test images. The Flickr2K dataset contains 2,650 high-resolution images with various contents and sizes. 

For preprocessing the training data in the DIV2K dataset, where the height ranges from 1116 to 2040 and the width ranges from 648 to 2040, we select images whose dimensions meet one of the following criteria: height $\geq$ 2040 and width $\geq$ 1152, or height $\geq$ 1152 and width $\geq$ 2040. We resize and rotate these images to a uniform size of 2040 × 1152 (height × width). This results in 704 images for the new training dataset, which we maintain as high-resolution (HR) and resize to 510 × 288 as low-resolution (LR) samples.

We apply the same preprocessing to the Flickr2K training set, where the height ranges from 1200 to 2040 and the width ranges from 1140 to 2040. Following the same criteria as the DIV2K dataset, we obtain 1986 images, and split them into training, validation, and test sets in the ratio of 0.8:0.1:0.1.
\item Test-bed:

We evaluate Sebica on a commodity computer equipped with standard hardware components: an Intel® Core i7-4790 CPU, 32 GB of RAM, and an NVIDIA GeForce GTX 2080Ti GPU, all running on Ubuntu 22.04 LTS. The implementations utilize Python 3.10, PyTorch 2.0.1, and OpenCV-python 4.7.0.

\item Training Process:

We use the training and validation datasets from Div2K and Flickr2K to train the network. Initially, we train the network on Div2K for 100 epochs, followed by 10 epochs on Flickr2K. For each dataset, we apply random cropping, rotations, and flips as augmentations, set the learning rate to \(1 \times 10^{-4}\), and use the optimizer \texttt{CosineAnnealingWarmRestarts} with a first cycle length of 1000 iterations and a minimum learning rate of \(1 \times 10^{-6}\). 
\end{itemize}

\subsection{Comparison with Sebica and Baseline}
We evaluate both Sebica and Sebica\_small in this section. Unless specifically stated otherwise, "Sebica" refers to the standard version when discussed below.

To demonstrate the superior performance, we compare our proposed network with the Bicubic Interpolation method and three other state-of-the-art (SOTA) algorithms that offer a good trade-off between high-resolution image reconstruction efficiency and quality. FSRCNN~\cite{dong2016accelerating} is a lightweight version of SRCNN optimized for real-time efficiency with fast inference on low-resolution inputs. EDSR~\cite{lim2017enhanced} leverages residual blocks to achieve significant high precision in high-quality images. RVSR, one of the winners in the AIS 2024 Challenge, a real-time 4K super-resolution challenge~\cite{conde2024real}, performs high-quality SR in real-time. These algorithms serve as our baselines for comparison.

\begin{figure}[htb]
\centering
\includegraphics[width=1.0\columnwidth]{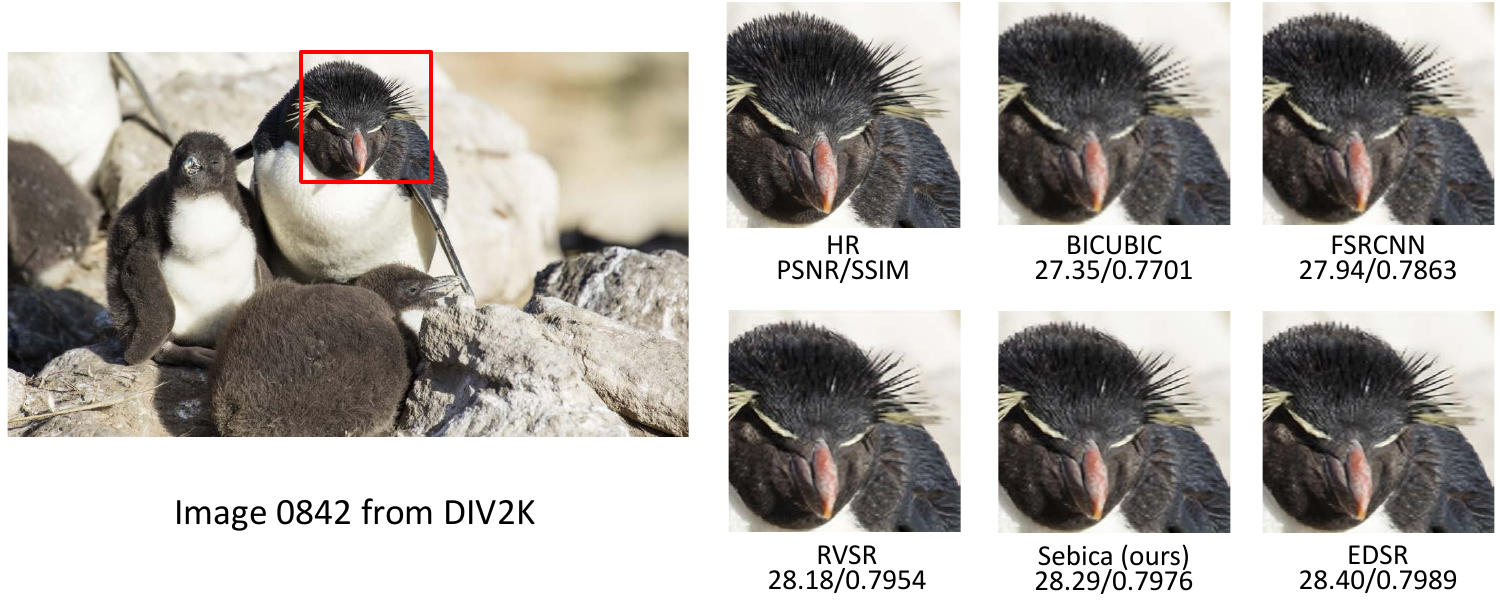}
\caption{\label{fig:comparison} 
Comparison of the SR images derived from baseline.
}
\end{figure}

In addition to Figure~\ref{fig:comparison_1}, Figure~\ref{fig:comparison} illustrates the quality of HR reconstruction compared to the baseline and ground truth (HR).

Furthermore, we provide detailed information regarding PSNR, SSIM metrics, as well as network size and computational complexity in terms of parameter size (K) and GFLPs, as shown in Table~\ref{tab:main_result}.

\begin{table}[htb]
\caption{Performance and model size comparison between our networks and selected methods. All networks listed here support 4 times upscaling of images to 1280×720. \textbf{Boldface}: best; \underline{Underlined}: second best.}
\setlength{\tabcolsep}{0.7 mm}
  \centering  
  \fontsize{8}{10}\selectfont  
  \begin{tabular}{ccccccc}
  \hline
Methods& \multicolumn{2}{c}{PSNR} & \multicolumn{2}{c}{SSIM} &  Params (K) &GFLOPs  \\ 
\cmidrule(lr){2-3} \cmidrule(l){4-5} 
& Div2K & Flirkr2K & Div2K & Flirkr2K &         &          \\
\hline\hline 
{Bicubic} &  27.35 & 29.70 & 0.7701&  0.8176 & - & -\\\hline
{FSRCNN} & 27.94 & 30.01 & 0.7863 & 0.8305 &\underline{12.0} & 5.00\\\hline
{EDSR} & \textbf{28.40} & \textbf{30.31} & \textbf{0.7989} & \textbf{0.8376} &241.0 & 14.15\\\hline
{RVSR} & 28.18 & 30.15 & 0.7954 & 0.8324 &221.1 & 10.87\\\hline
{Sebica} & \underline{28.29} & \underline{30.18} & \underline{0.7976} & \underline{0.8330} & 40.9 &\underline{2.10}\\\hline
{Sebica\_small} & 28.12 & 30.09 & 0.7931& 0.8317 & \textbf{7.9} &\textbf{0.41}\\\hline
\end{tabular}
\label{tab:main_result}
\end{table}
\label{sec:psrn_ssim}

We use DIV2K and Flirkr2K's test datasets, discussed in Section~\ref{sec:hw}, to evaluate the HR image reconstruction quality in terms of PSNR and SSIM. 

Table~\ref{tab:main_result} illustrates that Sebica significantly reduces computational costs while maintaining high reconstruction quality. It achieves PSNR/SSIM scores of 28.29/0.7976 and 30.18/0.8330 on the Div2K and Flickr2K datasets, respectively. These results outperform most baseline lightweight models and are comparable to the highest-performing model, which achieves PSNR/SSIM scores of 28.40/0.7989 and 30.31/0.8376 but with 22\% and 20\% more parameters and GFLOPs. Moreover, the small version of Sebica has only 7.9K parameters and 0.41 GFLOPS, which is just 3\% of the parameters and GFLOPs of the highest-performing model, while still achieving PSNR and SSIM metrics of 28.12/30.09 and 0.7931/0.8317, respectively, on the Div2K and Flickr2K dataset.

\subsection{Test in Object Detection Task}
Super Resolution (SR) is commonly employed in an edge-server system, where the edge device captures and transmits video as small-sized, i.e. low-resolution (LR) format, images to the server, and the SR process restores them to a higher resolution at the server's end. In this experiment, we compare the object detection performance, measured by mean Average Precision (mAP), between the original high-resolution video and the SR-restored video.

We use the dataset from \cite{bemposta2022dataset} to evaluate detection performance in terms of mAP by comparing LR images with and without SR, as well as against the original high-resolution (HR) images. This dataset contains 15,070 UAV-captured road traffic images across 12 sections, with 155,328 labeled vehicles (cars and motorcycles). For our testing, we utilize sec6, which records interurban traffic, comprising 692 images with a resolution of 720 x 1280, resized to 180 x 320 for the LR format.

We utilize YOLOV5s as the object detector and apply Sebica to reconstruct low-resolution (LR) images into high-resolution (HR) images, which are then fed into the detector. We evaluate performance using Mean Average Precision (mAP), as shown in Table~\ref{tab:mAP}. The results demonstrate that with Sebica's enhancement, the mAP increases from 0.446 to 0.487, 0.446 to 0.478, and 0.307 to 0.366 at IoU thresholds of 0.1, 0.2, and 0.5, respectively, closely matching the results achieved with ground-truth HR images. We also tested the small Sebica version and observed similar outcomes.

\begin{table}[htb]
\caption{Mean Average Precision (mAP)  of object detection with and without Sebica enhancement across various IoU thresholds.
}
\setlength{\tabcolsep}{4 mm}
  \centering  
  \fontsize{8}{10}\selectfont  
  \begin{tabular}{c|c|c|c}
  \hline  
IoU threshold & 0.1 & 0.2 & 0.5  \\\hline   \hline  
LR &  0.446 & 0.446 & 0.307 \\\hline
GT &  0.462 & 0.461 & \underline{0.363}\\\hline
Sebica  & \textbf{0.487} & \textbf{0.486} & \textbf{0.366}  \\\hline
Sebica\_small  & \underline{0.478} & \underline{0.478} & 0.347  \\\hline

\end{tabular}

\label{tab:mAP}
\end{table}

\section{Conclusions and Future Work}\label{sec:conc}
We propose a novel network called Sebica for Single Image Super-Resolution (SISR), which delivers high-quality results with low computational cost and without the use of compression techniques. This network utilizes spatial and efficient bidirectional channel attention, enhancing feature representation by adaptively recalibrating spatial and channel-wise information in both directions, thus improving both efficiency and accuracy. Compared to the baseline, Sebica significantly reduces computational cost while maintaining reconstruction quality. We also apply the network to an object detection task, showing that Sebica effectively reconstructs low-resolution images and improves detection accuracy.

In the future, we plan to expand and optimize the algorithm to process videos by handling consecutive frames and using attention mechanisms to capture object relationships across frames.

% \bibliographystyle{unsrt}
% \bibliography{ref.bbl}

\end{document}